\documentstyle[11pt,newpasp,twoside,epsf,epsfig]{article}
\def\edcomment#1{\iffalse\marginpar{\raggedright\sl#1\/}\else\relax\fi}
\reversemarginpar
\begin{document}

\def\etal{{\it et~al.~}}
\def\bsax{{\it BeppoSAX~}}
\def\ginga{{\it Ginga~}}
\def\astroe{{\it ASTRO-E~}}
\def\constellation{{\it CONSTELLATION~}}
\def\einstein{{\it Einstein~}}
\def\exosat{{\it EXOSAT~}}
\def\tenma{{\it Tenma~}}
\def\asca{{\it ASCA~}}
\def\rosat{{\it ROSAT~}}
\def\rxte{{\it RXTE~}}
\def\euve{{\it EUVE~}}
\def\heao1{{\it HEAO-1~}}
\def\integral{{\it INTEGRAL~}}
\def\cl{clusters of galaxies~}
\def\nt{nonthermal~}
\def\fov{field of view~}
\def\chandra{{\it Chandra~}}
\def\xmm{{\it XMM-Newton~}}
\def\kbeta{$K_{\beta}$~}
\def\kalpha{$K_{\alpha}$~}
\def\pho{~{\rm ph~cm}^{-2}~{\rm s}^{-1}~{\rm keV}^{-1}~}
\def\erg{~{\rm erg~ cm}^{-2}\ {\rm s}^{-1}~}
\def\ergs{~{\rm erg~s}^{-1}~}
\def\h0{~{\rm H_0 = 50~km~s}^{-1}\ {\rm Mpc}^{-1}~h_{50}~}

\title{First Results From an X-ray, Weak Lensing, and Sunyaev-Zeldovich Effect Survey of Nearby Clusters: Abell 3266}
\author{Percy L. G\'{o}mez}
\author{Kathy A. Romer, Jeff Peterson, Chris Cantalupo\altaffilmark{1}}
\affil{Carnegie Mellon University, Department of Physics, 5000 Forbes Avenue, Pittsburgh, USA} 
\author{Bill Holzapfel, Chao-Lin Kuo, Matt Newcomb}\affil{UC Berkeley, USA}
\author{John Ruhl\altaffilmark{2}, Jon Goldstein, Eric Torbet}\affil{UC Santa Barbara, USA}
\author{Marcus Runyan}\affil{Caltech, USA}

\altaffiltext{1}{Current affiliation: University of California
Berkeley Space Sciences Laboratory}
\altaffiltext{2}{Current affiliation: Case Western Reserve University}

\begin{abstract}

As part of a combined Sunyaev-Zeldovich Effect (SZE), X-ray and weak
lensing survey of low redshift ($z<0.1$) X-ray clusters, we present
SZE images of the $z=0.059$ X-ray cluster Abell 3266 at three
observing frequencies (150, 220, 275 GHz) and after the spectral
subtraction of primary Cosmic Microwave Background (CMB)
anisotropies. These images were generated using the ACBAR bolometer
array operated on the Viper telescope at the South Pole.  The
multi-frequency data from ACBAR should allow us to overcome one of the
main obstacles facing the analysis of SZE observations of nearby
clusters, {\it i.e.} contamination from primary Cosmic Microwave
Background (CMB) anisotropies.

\end{abstract}

\section{Introduction}

We are undertaking a multiwavelength study of the physical properties
of a sample of $\sim$ 10 nearby (z $<$ 0.1) X-ray bright galaxy
clusters. These clusters were selected from the REFLEX Survey of
southern X-ray clusters (B\"{o}hringer et al. 2001). They conform to
the following criteria: L$_x > 4\times 10^{44}$ erg/s [0.1-2.5 keV],
$z<0.1$ and $\delta < -44^{\circ}$.

In this study, we will combine new and archival X-ray observations
with weak lensing observations and new Sunyaev-Zeldovich Effect (SZE)
observations. The multiwavelength observations consist of X-ray
(archival ROSAT and new Chandra and XMM data), weak lensing (4m CTIO
observations already at hand), and new millimeter observations (from
the ACBAR bolometer array mounted on the Viper Telescope located at
the South Pole) intended to detect the SZE towards these clusters.

The Sunyaev-Zeldovich effect (Sunyaev \& Zel'dovich 1972) describes
the inverse Compton scattering of CMB photons by hot electrons in the
intra-cluster plasma.  There are two SZ effects: the kinetic and
thermal. The thermal effect is the larger by about an order of
magnitude.  The thermal effect is due to the random thermal motion of
the cluster electrons and shifts lower energy CMB photons to higher
energy, while the kinetic effect is due to the bulk motion of the
cluster. Interestingly, the thermal SZE spectrum has a null at about
218 GHz where the kinetic SZE is maximum.

SZE measurements can be performed with three different techniques:
interferometers, single-dish radiometric observations and bolometers
(for a recent review see Birkinshaw 1999).  Interferometer studies have
produced high signal-to-noise SZ maps of roughly 20 clusters ({\it
e.g.}, Carlstrom et al. 1996; Reese et a. 2002; Jones et al. 2001;
Udomprasert, Mason \& Readhead 2001).  These maps take advantage of
the spatial filtering and high stability intrinsic to interferometer
observations. However, interferometers have been used only at low
frequencies (well below the thermal null) and suffer of variable radio
source contamination. Single dish radiometric observations have also
been performed ({\it e.g.} Birkinshaw, Hughes \& Arnaud 1991;
Cantalupo et al. 2002), but these are also restricted to frequencies
below the thermal null. On the other hand, bolometers offer high
sensitivity and an extended spectral range. They operate from a few to
several hundred GHz and so can gather data below, at and above the
null in the thermal SZE ({\it e.g.}, Holzapfel et al. 1997;
Pointecouteau et al.2001).

The rationale behind combining multiwavelength observations of
clusters is that they probe the cluster components in different
ways. For instance, X-ray and SZE data probe the gas density and
temperature distributions, whereas weak lensing observations probe the
total mass distribution. The combination of X-ray and SZE data
provides three observables (X-ray surface brightness, SZE decrement,
and projected temperature) which constrain the variation of density
and temperature throughout the target cluster. Comparing weak-lensing
mass distributions with X-ray and SZE gas distributions in the same
clusters allows a detailed test of usual assumptions (hydrostatic
equilibrium, spherical symmetry, and the lack of significant
substructure) on which estimates for the amount and distribution of
dark matter in clusters are based.

Our study will attempt to: 1) characterize the density, temperature,
and dynamical state of the gas in each cluster. 2) compare the mass
estimates derived from the SZE and X-ray observations with the more
direct weak lensing mass. This study of nearby clusters is paramount
to determine whether, and to what level, future cosmological parameter
estimates from SZE and X-ray cluster surveys might be biased by
non-thermal cluster physics and/or by non standard cluster geometries
(e.g., Holder et al. 2001). Here we highlight observations of one of
the brightest members of our sample, A3266.

\section{Abell 3266}

Abell 3266 has an X-ray temperature of 6.2 keV (Markevitch et
al. 1998) and a redshift of $z=0.059$.  There is evidence that
suggests that this cluster is a postmerger system (Quintana et
al. 1996, Mohr et al. 1993). Recently, Roettiger \& Flores (2000) have
modeled this cluster as the result of an off-axis cluster
merger. They found that a 2.5:1 mass ratio off-axis merger provides a
good match to the ROSAT temperature and gas density distribution for
the clusters. In addition, they have also predicted the presence of
large rotating bulk flows in this cluster.

\subsection{Weak Lensing and X-ray Data}

A3266 was observed by the CTIO 4m telescope in the R and I band in
1998 in order to create a weak lensing map (Joffre et al. 2001). These
observations covered a FOV of $45'\times 45'$ and achieved $\sim$ 28.4
mag/arcsec$^2$ and 26.9 mag/arcsec$^2$ in the R and I bands
respectively. Overall, some 36000 galaxies ellipticities were measured
in the field of this cluster (Joffre et al. 2001). Joffre et
al. pointed out that the mass is concentrated towards the cD galaxy.

We have retrieved and processed archival ROSAT pointed X-ray
observations of Abell 3266. We have used Snowden et al (1994) ESAS
process to create count maps in the 0.4-2.0 keV band. In order to
deproject the density distribution, we perform a fully tridimensional
projected beta model maximum likelihood fit to the ROSAT data
(Cantalupo et al. 2002) which yields a $\beta$ value of 0.92 $\pm$
0.02 and a core radius of 423 $\pm$ 20 kpc.

\subsection{Multifrequency Sunyaev Zeldovich Effect Observations}

ACBAR (Runyan et al. 2002) is a 16 pixel multifrequency mm-wave
bolometer array and operates on the 2m Viper telescope at the South
Pole. During 2001, ACBAR operated simultaneously at 150, 220, 275, and
350GHz, {{\it i.e.} on either side of the thermal null, and -- because
of the superb transmission and stability of the South Pole atmosphere
-- achieved high sensitivity to clusters.  The ACBAR beam size is
roughly $5'$ (FWHM) at each frequency. Typical sensitivities for each
150GHz detector are 400 $\mu{\rm{K{}_{CMB}s^{1/2}}}$.

We observed Abell 3266 in March and in August of 2001 and were able to
create maps covering roughly one square degree centered on the cluster
at each of the following frequencies: 150, 220, 275 GHz. The typical
random noise in these maps was 30 $\mu $K per beam at 150. As expected
these maps were contaminated by primary CMB anisotropies. This
contamination was especially pronounced in the 150 map, where the CMB
noise dominated the random noise.  However, our multifrequency
observations and our knowledge of the spectrum of the physical
mechanism responsible for the thermal SZE allowed us to spectrally
filter out the CMB primary anisotropy.  Intuitively, this method is
very simple as it can be shown that at 150 GHz we detect the CMB minus
the cluster signal; at 220 GHz -the thermal SZE null- the CMB
dominates the map; and at 275 GHz we detect the CMB plus the cluster
signal. So, naively subtracting the 150 GHz data from the 275 GHZ data
would retrieve the cluster signal and cancel out the CMB data. In
practice, we construct a linear combination of the 150, 220, and 275
GHz images designed to maximize the SZ signal and minimize the CMB
signal at the RJ limit that takes into account the shape of the SZE
spectrum.  Figure 1 shows that ACBAR can use the SZE to resolve some
of the same extended structures seen in the ROSAT X-ray image and that
most of the primary CMB anisotropy signal has been subtracted in the
CMB subtracted map.

\section{Conclusions}

In this preliminary analysis of Abell 3266, we have demonstrated that
we can use the three ACBAR observing frequencies to remove CMB signal
in maps of the SZE. In the future, we expect to improve the SZE data
processing and to obtain SZE maps and SZE spectra for all the clusters
in our sample.  Through the combination of ACBAR, weak lensing and
X-ray data, we should obtain a better understanding of cluster physics
and hence of the role of clusters as cosmological probes.

\acknowledgements

This research was supported by NASA grant number NAG5-7926. We thank
P. Ade, A. Lange, M. Daub, M. Leuker, and J. Bock for their
contributions to ACBAR; R. Nichol, T. McKay, and J. Freemann for
allowing us access to the weak lensing data; and K. Bandura for
assistance with the X-ray fitting.

\begin{figure}
\plotone{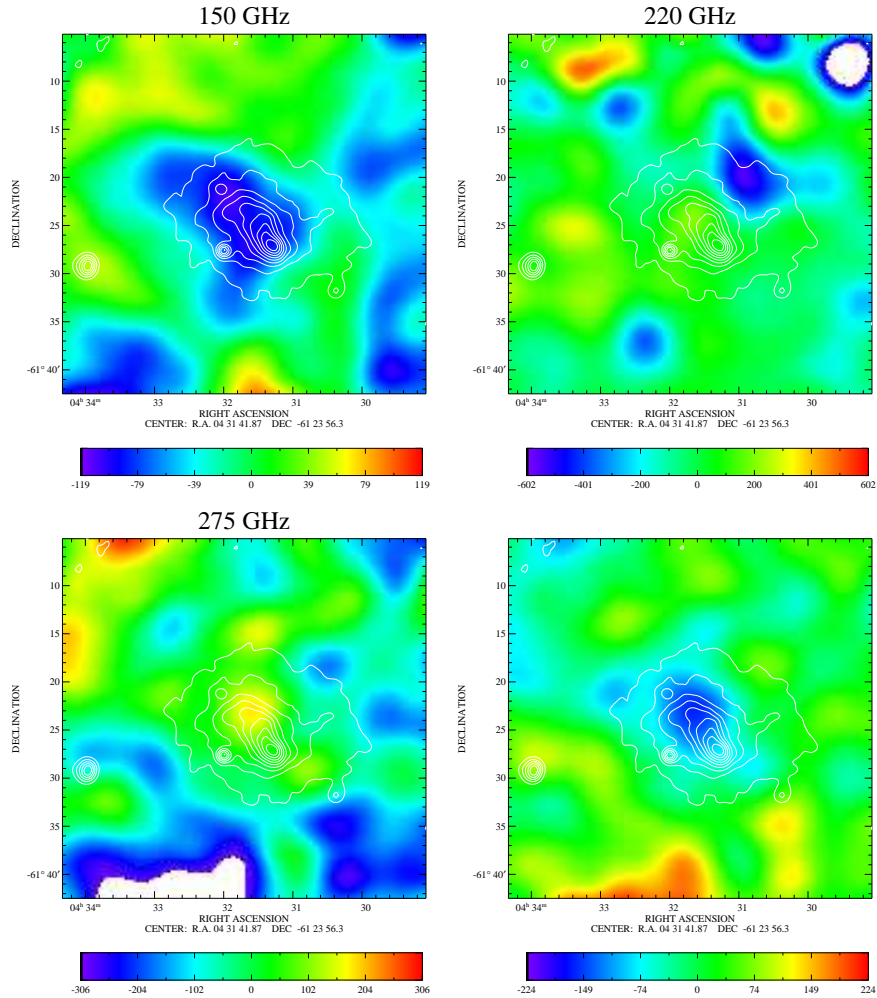}
\vspace{-1in}

\caption{\small Mosaic of the 150 GHz, 220 GHz, 275 GHz, and  CMB 
spectrally subtracted colorscale images of Abell 3266 (convolved with
a Gaussian with FWHM $\sim 5'$ overlaid onto the ROSAT contours.
Note that most of the CMB present in the 150 GHz, 220GHz, and 275GHz
channels has been minimized in the CMB subtracted map and that the
strongest signal is associated with the cluster. As expected, the SZE
cluster signal at 220GHz is minimum.  }

\end{figure}

\end{document}